\documentclass[pra,showpacs,superscriptaddress]{revtex4-2}
\usepackage{amsmath}
\usepackage{amsfonts}
\usepackage{graphicx}
\usepackage{longtable}
\newcommand{\be}{\begin{equation}}
\newcommand{\ee}{\end{equation}}

\newcommand{\ba}[1]{\left(\begin{array}{#1}}
\newcommand{\ea}{\end{array}\right)}
\begin{document}
\title{Local sum uncertainty relations for angular momentum operators of bipartite permutation symmetric systems}
\author{I. Reena} 
 \affiliation{Department of Physics, Jnanabharathi, Bangalore University, Bangalore-560056, India}
\author{H. S. Karthik} 
\affiliation{International Centre for Theory of Quantum Technologies, University of Gdansk, Gdansk, 80-308,  Poland}
\author{J. Prabhu Tej}
\affiliation{Department of Physics, Ramaiah University of Applied Sciences, Bangalore-560054}
\author{A. R. Usha Devi} 
\email{arutth@rediffmail.com}
\affiliation{Department of Physics, Jnanabharathi, Bangalore University, Bangalore-560056, India}
\affiliation{Inspire Institute Inc., Alexandria, Virginia, 22303, USA.}
\author{Sudha} 
\affiliation{Department of Physics, Kuvempu University, 
	Shankaraghatta-577 451, India}
\affiliation{Inspire Institute Inc., Alexandria, Virginia, 22303, USA.}
\author{A. K. Rajagopal} 
\affiliation{Inspire Institute Inc., Alexandria, Virginia, 22303, USA.}
\date{\today}
\begin{abstract} 
We show that violation of variance based local sum uncertainty relation (LSUR) for angular momentum operators of a bipartite system,  proposed by Hofmann and Takeuchi~[Phys.Rev.A {\bf 68}, 032103 (2003)], reflects entanglement in the equal bipartitions of an $N$-qubit symmetric state with even qubits. We establish the one-to-one connection with the violation of LSUR with negativity of covariance matrix [Phys. Lett. A {\bf 364}, 203 (2007)] of the two-qubit reduced system  of a permutation symmetric $N$-qubit state.  
\end{abstract}
\maketitle 
\section{Introduction} 
Uncertainty relations place fundamental limits on the precision achievable in  measuring non-commuting observables. The original idea of uncertainty relation was first introduced  by  Heisenberg~\cite{heis}  for position and momentum observables $Q$, $P$. Subsequently,  a  mathematically precise version (with $\hbar=1$)         
\be
\label{pqunc}
(\Delta Q)^2 \, (\Delta P)^2 \geq \frac{1}{4}
\ee
was  formulated by Kennard~\cite{kennard}. Here  $(\Delta \Gamma)^2 = \langle \Gamma^2  \rangle - \langle \Gamma  \rangle^2$ denotes the variance of the observable $\Gamma=Q \ {\rm or}\  P$ and  the bracket $\langle \cdots \rangle=\mbox{Tr}\,[\rho\cdots ]$ corresponds to the expectation value in a quantum state $\rho$. 

Weyl~\cite{weyl} and Robertson~\cite{robt} extended the uncertainty relation  (\ref{pqunc}) for any arbitrary pair of physical observables $A_1$, $A_2$:    
\be
\label{unc}
(\Delta A_1)^2 (\Delta A_2)^2 \geq \frac{1}{4}\left\vert 
\langle\,[A_1,\,A_2 ]\,\rangle \right\vert^2
\ee
which is commonly referred to as the Heisenberg-Robertson uncertainty relation in the literature. Here $[A_1,\,A_2 ]=A_1A_2-A_2A_1$ denotes the commutator of the observables $A_1$ and $A_2$. The uncertainty relation (\ref{unc}) imposes restrictions on the product of variances 
$(\Delta A_1)^2$ and $(\Delta A_2)^2$  --   essentially limiting the capability towards precise prediction of the  measurement results of non-commuting observables. In general the Heisenberg-Robertson approach of placing limits on the uncertainties of incompatible observables in the given quantum state $\rho$  sets the conventional framework for deriving preparation uncertainty relations$^{1}${\footnote{$^{1}$Preparation uncertainty relations place intrinsic bounds on the spread of incompatible observables, measured in different statistical trials, using  identical preparations of a quantum state.  On the other hand, measurement uncertainty relations impose bounds on the  {\em error} occuring in the measurement scheme of one of the observables and the corresponding post-measurement {\em disturbance} caused on the other observable in an apparatus.}}~\cite{busch_werner}.

Apart from their fundamental interest, uncertainty relations play a significant role in the field of quantum information processing, with several applications such as entanglement detection~\cite{hf,hfbound, ogunhe},  quantum cryptography~\cite{koshi, berta, hangi, tom, branciard, aru_hsk, coles}, quantum metrology~\cite{reid2011}  and   foundational tests  of quantum theory~\cite{scully,reid2011}. Motivated by their applicability,  there has  been an ongoing interest in  reformulating uncertainty relations expressing trade-off of more than two incompatible observables, formalized in terms of  variances~\cite{hf,ogunhe,akp1,rivas,varbase,akp2,chen,shabbir,xiao,bagchi,ma,song,bsanders,maconne19, busch2019,zheng,zukow}, or via information entropies~\cite{ hirschman, beckner, bial1, deutch, partovi, bial2, kraus, mu, sw, bial3}. Recently several experimental tests have been carried out to verify different forms of uncertainty relations~\cite{expt_wang, expt2017, chen_expt, expt2019}.

In this paper we investigate variance based LSUR for local angular momentum operators of a bipartite quantum system,  proposed  by Hofmann and Takeuchi~\cite{hf} and examine the implications of its violation in bipartite permutation symmetric systems. We show that the angular momentum  LSUR, which places lower bound on the set of all bipartite separable states,  gets  violated if and only if the covariance matrix of the two-qubit reduced system of the $N$-qubit permutation symmetric state is not positive semi-definite. Since it has been shown ~\cite{ijmp06,pla07,prl07} that the covariance matrix negativity serves as a necessary and sufficient condition for entanglement in a two-qubit symmetric system, our result establishes a one-to-one connection between violation of the LSUR and pairwise entanglement.

\section{Sum Uncertainty Relation}

It may be noted that the term  $\langle\,[A_1,\,A_2 ]\,\rangle={\rm Tr}\left(\rho\,[A_1,\,A_2] \right) $ appearing in the right hand side of the Heisenberg-Robertson uncertainty relation  
 (\ref{unc})  vanishes in some specific quantum states $\rho$.  In such cases, one ends up with a  trivial relation $(\Delta A_1)^2 (\Delta A_2)^2 \geq 0$ for the product of variances of non-commuting observables $A_1, \, A_2$.  
 Moreover,  variance   vanishes in the eigenstate of one of the observables. In such cases the Heisenberg-Robertson uncertainty relation (\ref{unc}) fails to capture the intrinsic indeterminacy of non-commuting observables.  To overcome such issues it is preferable to  employ uncertainty relations placing non-trivial bounds on the sum of variances $(\Delta A_1)^2 + (\Delta A_2)^2$. In fact, a lower bound for the sum of variances may be found  by using  the inequality 
 $\sum_{\alpha=1}^{m} a_\alpha/m \geq  \left(\prod_{\alpha}a_\alpha\right)^{1/m}$ between the arithmetic mean and the geometirc mean of real non-negative numbers $a_\alpha, \alpha=1,2,\ldots, m$. Choosing $a_1=(\Delta A_1)^2,\  a_2=(\Delta A_2)^2$ and using (\ref{unc}) one obtains a variance based sum uncertainty relation  
\be
\label{sur0}
(\Delta A_1)^2 + (\Delta A_2)^2\geq \left\vert \langle\,[A_1,\,A_2 ]\,\rangle \right\vert.
\ee

However the sum uncertainty relation (\ref{sur0}) is a byproduct of the Heisenberg-Robertson inequality (\ref{unc}) and thus it is  non-informative for some quantum states $\rho$ in which one of the variances and/or  $\langle\,[A_1,\,A_2 ]\,\rangle$ vanish.  

\subsection{Sum uncertainty relations for angular momentum operators}

Hofmann and Takeuchi~\cite{hf}  proposed that  a non-trivial lower bound ${\cal U}>0$ must exist for the sum of variances $\sum_{\alpha}\,(\Delta A_\alpha)^2$ of a set $\{A_\alpha\}$, $\alpha=1,2,\ldots $ of    
non-commuting observables, as they do not share any simultaneous eigenstate i.e., for non-commuting observables $\{A_\alpha\}$, $\alpha=1,2,\ldots$, the following inequality holds good~\cite{hf}:
\be 
\label{sur1}
\sum_{\alpha}\,(\Delta A_\alpha)^2\geq {\cal U}.
\ee
The lower bound ${\cal U}$ in the sum uncertainty relation (\ref{sur1}) corresponds to the absolute  minimum value $\left[\sum_{\alpha}\, (\Delta A_\alpha)^2\right]_{\rm min}$ for any quantum state $\rho$. While it is tough to determine  ${\cal U}$ for any arbitrary set  $\{A_{\alpha}\}$ of non-commuting observables in general, there are some important physical examples  in finite dimensional Hilbert spaces, where the limiting value ${\cal U}$ can be readily identified. For example, in the case of a spin-$j$ quantum system the components of angular momentum operators 
$\{J_1, J_2, J_3\}$ satisfy the following conditions: 
\begin{eqnarray*}
 \left\langle \left(J^2_1+J^2_2+J^2_3\right)\right\rangle&=&j(j+1), \ \ \ \ \ \langle J_1\rangle^2+\langle J_2\rangle^2+\langle J_3\rangle^2 \leq j^2. 
  \end{eqnarray*}
Thus, one obtains a variance based  sum uncertainty relation for the components $(J_1, J_2, J_3)$ of angular momentum operator~\cite{hf}: 
\be
\label{sur} 
(\Delta J_1)^2+(\Delta J_2)^2+(\Delta J_3)^2 \geq j,  
\ee
imposing a limit on the measurement precision of more than one of the angular momentum components.
 
In the specific  example of $j=1/2$ (i.e., for a qubit), we have $J_\alpha=\sigma_\alpha/2, \alpha=1,2,3$, where $\sigma_\alpha$ denote the Pauli matrices.  One then obtains the sum uncertainty relation for qubits~\cite{hf} 
\be
\label{paulisur}
(\Delta \sigma_1)^2+(\Delta \sigma_2)^2+(\Delta \sigma_3)^2 \geq 2.
\ee  
Given an arbitrary single qubit density matrix $\rho$, expressed in the standard basis $\{\vert 0\rangle, \vert 1\rangle\}$, 
\begin{eqnarray}
\rho&=&\frac{1}{2}\left(I+\sum_{\alpha=1}^3\,  s_\alpha \sigma_\alpha\right), \ \ \ \alpha={\rm Tr}[\rho\, \sigma_\alpha]; \ \ \ \sum_{\alpha} s_\alpha^2\leq 1,  
\end{eqnarray}
one obtains $(\Delta \sigma_1)^2+(\Delta \sigma_2)^2+(\Delta \sigma_3)^2=3-\sum_{\alpha} s_\alpha^2\geq 2$, in accordance with the sum uncertainty relation (\ref{paulisur}). Equality sign holds when $\sum_\alpha\, s_\alpha^2=1$ i.e., for pure states of qubit.     

\subsection{Local sum uncertainty relations for bipartite systems}

Let us consider angular momentum operators $J_{A\alpha}, \ J_{B\alpha}$ acting on the Hilbert spaces ${\cal H}_A$, ${\cal H}_B$ of dimensions $d_A=(2j_A+1)$, $d_B=(2j_B+1)$ respectively. They satisfy the sum uncertainty relations 
\be
\label{main}
\sum_{\alpha=1}^3\, (\Delta J_{A\alpha})^2 \geq j_A, \ \  \sum_{\alpha=1}^3\, (\Delta J_{B\alpha})^2 \geq j_B. 
\ee   
Then,  the  LSUR~\cite{hf} 
\begin{eqnarray}
\label{lsur}
\sum_{\alpha=1}^3\, \left[\Delta \left(J_{A\alpha}+J_{B\alpha}\right)\right]^2\geq j_A+j_B
\end{eqnarray}
is necessarily satisfied by the set of all bipartite separable states $\rho^{(\rm sep)}_{AB}=\sum_{k}\, p_k\,\left(\rho_{Ak}\otimes \rho_{Bk}\right)$ in the Hilbert space ${\cal H}_A\otimes {\cal H}_B$.  Violation of the LSUR (\ref{lsur})  serves as  a  clear signature of entanglement. This is readily seen by considering a bipartite system with $j_A=j_B=j$ prepared in a  spin singlet state 
$$\vert\Psi^{\rm singlet}_{AB}\rangle=\frac{1}{\sqrt{2j+1}}\sum_{m=-j}^j\, (-1)^{j-m}\, \vert j, m\rangle_A \otimes \vert j,- m\rangle_B$$ 
in which one obtains  
$$\left(J_{A\alpha}+J_{B\alpha}\right)\vert\Psi^{\rm singlet}_{AB}\rangle\equiv 0,\ \  \alpha=1,\,2,\,3,$$ leading to $$\sum_{\alpha=1}^3\, \left[\Delta \left(J_{A\alpha}+J_{B\alpha}\right)\right]^2=0.$$  In other words, the LSUR (\ref{lsur}) gets violated in  a  spin singlet state $\vert\Psi^{\rm singlet}_{AB}\rangle$, thus highlighting  the entanglement property that local observables  of the subsystem $A$ can be   determined by performing measurements on the other subsystem $B$. More specifically, violation of LSUR signifies in general that correlations between subsystems in an entangled state can be determined with enhanced precision than those in a separable state.           

It is  of interest to explore if violation of local sum uncertainty relations is both necessary and sufficient to detect  entanglement in some special classes of bipartite quantum systems. To this end, Hoffmann and Takeuchi~\cite{hf} considered the following two-qubit LSUR 
\be
\label{lsur_qubit}
\sum_{\alpha=1}^3\,\left[\,\Delta(\sigma_\alpha \otimes I+I \otimes \sigma_\alpha)\right]^2\geq 4,
\ee
obtained by substituting $j_A=j_B=1/2$ and $J_{A\alpha}=(\sigma_\alpha\otimes I)/2$, $J_{B\alpha}=(I\otimes \sigma_\alpha)/2,\, \alpha=1,\, 2,\, 3$ in (\ref{lsur}). 
In the one-parameter family of two-qubit Werner states, 
\be
\label{werner}
\rho^{\rm Werner}_{AB}=\frac{(1-x)}{4}(I_2\otimes I_2)+x\, \vert\Psi\rangle_{AB}\langle \Psi\vert,  
\ee 
where $\vert\Psi\rangle_{AB}=\frac{1}{\sqrt{2}}\, (\vert 0_A\,1_B\rangle-\vert 1_A\,0_B\rangle$ and $0\leq x\leq 1$. It is readily seen that 
 $$[\Delta(\sigma_\alpha \otimes I+I \otimes \sigma_\alpha)]^2=2\, (1-x), \ \ \alpha=1,2,3.$$ 
 Thus, the left hand side of LSUR (\ref{lsur_qubit}) is given by              
\be
\label{WerLSUR}
\sum_{\alpha=1}^3\,\left[\,\Delta(\sigma_\alpha \otimes I+I \otimes \sigma_\alpha)\right]^2=6\, (1-x).
\ee
in the Werner class of two-qubit states, which are known to be entangled for $x> 1/3$.  From (\ref{WerLSUR}) it is evident that the LSUR (\ref{lsur_qubit}) is violated in the parameter range  $1/3<x\leq 1$. Thus,  violation of the two-qubit LSUR (\ref{lsur_qubit}) is both necessary and sufficient for detecting entanglement in the one-parameter family of two-qubit Werner states. 

In the next section  we discuss violation of LSUR  in permutation symmetric $N$-qubit states. 

\section{Violation of LSUR by permutation symmetric $N$-qubit states}
\label{sec2I}
Permutation symmetric $N$-qubit states draw  attention due to their experimental feasibility and  for the mathematical simplicity offered by them~\cite{ijmp06,pla07,prl07,sym01,sym05,aups,sym06,braun,markham,arus,lamata,symnew}. This class of states belongs to the $d=2j+1=N+1$ dimensional subspace of the $2^N$ dimensional Hilbert space, which  corresponds to the maximum value $j=N/2$ of angular momentum of $N$-qubit system. Invariance under exchange of particles labeled by $\alpha$, $\beta$ in a multiparty state $\rho^{\rm sym}$ is exhibited by the property   
\be 
\Pi_{\alpha\beta}\rho^{\rm sym}=\rho^{\rm sym}\Pi_{\alpha\beta}=\rho^{\rm sym}
\ee
where  $ \Pi_{\alpha\beta}$ denotes the permutation operator corresponding to the interchange of labels $\alpha,\, \beta$. 

We focus our attention on permutation symmetric $N$-qubit states ${\rho_{AB}^{\rm sym}}$ with even $N$, i.e., $N=2n,\ n=$integer (which corresponds to the value $j=N/2=n$ associated with  the set $\{J^{\rm total}_{\alpha}=J_\alpha\otimes I_2^{\otimes n}+I_2^{\otimes n}\otimes J_\alpha,\ \alpha=1,2,3\}$ of collective angular momentum operators  of the $N$-qubit system) and explore the LSUR  (\ref{lsur}) for bipartite divisions $A$, $B$  characterized respectively by  $j_A=j_B=n/2$. The angular momentum operators of the bipartitions $A$ and $B$ are explicitly given by    
\begin{eqnarray} 
\label{jajb}
J_{A\alpha}=J_\alpha\otimes I_2^{\otimes n}& =&\frac{1}{2}\, \sum_{k=1}^{n}\,\sigma_{k\alpha}\otimes I_2^{\otimes n}  \nonumber \\ 
J_{B\alpha}= I_2^{\otimes n} \otimes J_\alpha &=& I_2^{\otimes n}\otimes \frac{1}{2}\,\sum_{k=1}^{n}\,  \sigma_{k\alpha},  
\end{eqnarray}
where $I_2^{\otimes n}=I_2\otimes I_2\otimes\ldots \otimes I_2$ denotes $n$-th order tensor product of $2\times 2$ identity matrix  $I_2$  and 
\begin{equation}
\label{sigman}
\sigma_{k\alpha}=I_2\otimes I_2\otimes\ldots\otimes \sigma_\alpha\otimes I_2\otimes \ldots \otimes I_2
\end{equation}
with $\sigma_\alpha,\ \alpha=1,2,3$ appearing at the position $k\leq n$ in the $n^{\rm th}$ order tensor product. Using angular momentum algebra it readily follows that    
\begin{eqnarray}
\label{nnp1A}
\sum_{\alpha=1}^3\, \left\langle\, J^2_{A\alpha}\right\rangle_{\rho_{AB}^{\rm sym}}=j_A(j_A+1)=\frac{n(n+2)}{4} \\  
\label{nnp1B}
\sum_{\alpha=1}^3\, \left\langle\, J^2_{B\alpha}\right\rangle_{\rho_{AB}^{\rm sym}}=j_B(j_B+1)=\frac{n(n+2)}{4}. 
\end{eqnarray} 
Furthermore, the expectation values $\left\langle J_{A\alpha}\right\rangle_{\rho_{AB}^{\rm sym}}$, $\left\langle  J_{B\alpha}\,\right\rangle_{\rho_{AB}^{\rm sym}}$, $\left\langle J_{A\alpha}\,J_{B\alpha}\right\rangle_{\rho_{AB}^{\rm sym}}$ evaluated in the $N$-qubit symmetric state
 ${\rho_{AB}^{\rm sym}}$ can be expressed in terms of two-qubit averages as follows:    
\begin{eqnarray}
\label{s1}
\left\langle J_{A\alpha}\right\rangle_{\rho_{AB}^{\rm sym}}&=& \left\langle J_{\alpha}\otimes\,I_2^{\otimes n}\right\rangle_{\rho_{AB}^{\rm sym}}=\frac{1}{2}\, \sum_{k=1}^n\, \left\langle\,\sigma_{k\,\alpha} \otimes I_2^{\otimes{n}}\,\right\rangle_{\rho_{AB}^{\rm sym}}\nonumber  \\ 
&=&\frac{n}{2}\,  \langle\,\sigma_{\alpha}\otimes I_2 \rangle_{\varrho^{\rm sym}} \\
\left\langle  J_{B\alpha}\,\right\rangle_{\rho_{AB}^{\rm sym}}&=& \left\langle I_2^{\otimes n}\otimes J_{\alpha}\,\right\rangle_{\rho_{AB}^{\rm sym}}=\frac{1}{2}\, \sum_{l=1}^n\,  \left\langle\,  I_2^{\otimes{n}}\, \otimes \sigma_{l\,\alpha}  \right\rangle_{\rho_{AB}^{\rm sym}}\nonumber \\
\label{s2}
&=&\frac{n}{2}\,  \langle\,I_2\otimes\sigma_{\alpha}\rangle_{\varrho^{\rm sym}},\\ 
\label{scor}
\left\langle J_{A\alpha}\,J_{B\alpha}\right\rangle_{\rho_{AB}^{\rm sym}}&=& \left\langle J_{\alpha}\,\otimes J_{\beta}\right\rangle_{\rho_{AB}^{\rm sym}}=  \frac{1}{4}\, \sum_{k,l=1}^n\, \left\langle\,\sigma_{k\,\alpha} \otimes\,  \sigma_{l\,\alpha}\right\rangle_{\rho_{AB}^{\rm sym}},  \nonumber \\ 
&=&  \frac{n^2}{4}\,  \left\langle\,\sigma_{\alpha} \otimes\,  \sigma_{\alpha}\right\rangle_{\varrho^{\rm sym}}.
\end{eqnarray} 
Here it may be noticed  that    
\begin{eqnarray*}
 \left\langle\,\sigma_{k\,\alpha} \otimes I_2^{\otimes{n}}\,\right\rangle_{\rho_{AB}^{\rm sym}}&=& \langle\,\sigma_{\alpha}\otimes I_2 \rangle_{\varrho^{\rm sym}} \\ 
 \left\langle\,  I_2^{\otimes{n}}\, \otimes \sigma_{l\,\alpha}  \right\rangle_{\rho_{AB}^{\rm sym}}&=& \langle\,I_2\otimes\sigma_{\alpha}\rangle_{\varrho^{\rm sym}}, \\
\left\langle\,\sigma_{k\,\alpha} \otimes\,  \sigma_{l\,\alpha}\right\rangle_{\rho_{AB}^{\rm sym}}&=&\left\langle\,\sigma_{\alpha} \otimes\,  \sigma_{\alpha}\right\rangle_{\varrho^{\rm sym}}
\end{eqnarray*}
irrespective of the qubit labels $k,l=1,2,\ldots n$ (because the system is symmetric under interchange of constituent qubits)~\cite{ijmp06,pla07, prl07}, with   $\varrho^{\rm sym}$ denoting  the two-qubit reduced state of {\em any random pair} $(k,l)$ of qubits,  drawn from the $n$-qubit partitions  $A$ and $B$ of the  $N=2n$-qubit state ${\rho_{AB}^{\rm sym}}$. 

Expressing the two-qubit symmetric density matrix     
\begin{eqnarray}
\label{varrho}
\varrho^{\rm sym}&=&\frac{1}{4}\left(I_2\otimes I_2+\sum_{\alpha=1}^3 \, (\sigma_\alpha\otimes I_2+I_2\otimes \sigma_\alpha)\, s_\alpha+\sum_{\alpha,\beta=1}^3 \, (\sigma_\alpha\otimes \sigma_\beta)\, t_{\alpha\,\beta}\right) 
\end{eqnarray}    
in terms of its 8 real state parameters~\cite{ijmp06,pla07,prl07}
\begin{eqnarray}
\label{salpha0}
s_\alpha&=&\langle\,\sigma_{\alpha}\otimes I_2 \rangle_{\varrho^{\rm sym}}=\langle\,I_2\otimes\sigma_{\alpha}\rangle_{\varrho^{\rm sym}}  \\ 
\label{tab0}
t_{\alpha\,\beta}&=&\left\langle\,\sigma_{\alpha} \otimes\,  \sigma_{\beta}\right\rangle_{\varrho^{\rm sym}}=\left\langle\,\sigma_{\beta} \otimes\,  \sigma_{\alpha}\right\rangle_{\varrho^{\rm sym}} =t_{\beta\,\alpha},  \ \ \ \ \ \  t_{11}+t_{22}+t_{33}=1,  
\end{eqnarray}    
leads to the following identifications for the expectation values $\left\langle J_{A\alpha}\right\rangle_{\rho_{AB}^{\rm sym}}$, $\left\langle  J_{B\alpha}\,\right\rangle_{\rho_{AB}^{\rm sym}}$, $\left\langle J_{A\alpha}\,J_{B\alpha}\right\rangle_{\rho_{AB}^{\rm sym}}$ (see (\ref{s1}), (\ref{s2}), (\ref{scor})):
\begin{eqnarray}
\label{salpha1}
\left\langle J_{A\alpha}\right\rangle_{\rho_{AB}^{\rm sym}}&=&\frac{n}{2}\, s_\alpha
= \left\langle  J_{B\alpha}\,\right\rangle_{\rho_{AB}^{\rm sym}} \\   
\label{t}
\left\langle J_{A\alpha}\,J_{B\alpha}\right\rangle_{\rho_{AB}^{\rm sym}}&=& \frac{n^2}{4}\, t_{\alpha\alpha}
\end{eqnarray}

It may be noted that under {\em identical} local unitary transformation    $\varrho^{\rm sym}\rightarrow U^{\otimes 2}\varrho^{\rm sym}\left(U^{\dag}\right)^{\otimes 2}$,\   $U\in SU(2)$, the qubit orientation vector $s=(s_1,s_2,s_3)^T$ (see (\ref{salpha1})) and the $3\times 3$ real symmetric two-qubit correlation matrix $T=(t_{\alpha\,\beta}),\ \alpha,\beta=1,2,3$ (see (\ref{t})) transform  as~\cite{pla07}  
\begin{eqnarray}
 s\rightarrow s'&=&R\, s,\ \ \ \ \  T\rightarrow T'= R\,T\,R^T
\end{eqnarray}
where $R\in SO(3)$ denotes the  $3\times 3$ rotation matrix. 
  
Let us consider the angular momentum operators of the partitions $A$ and $B$ of the even $N$-qubit density matrix $\rho_{AB}^{\rm sym}$  (See (\ref{jajb})) 
\[ 
J_{A\alpha}=J_\alpha\otimes I_2^{\otimes n} =\frac{1}{2}\, \sum_{k=1}^{n}\,\sigma_{k\alpha}\otimes I_2^{\otimes n}, 
\]
\begin{eqnarray}
\label{jb}
J'_{B\alpha}&=&I_2^{\otimes n} \otimes \left\{ \left[U^\dagger(\hat{a},\theta)\,\right]^{\otimes n}\,J_\alpha \left[U(\hat{a},\theta)\right]^{\otimes n}\right\} \nonumber \\ 
&=& I_2^{\otimes n}\otimes \frac{1}{2}\,\sum_{k=1}^{n}\, \left\{\left[U^\dagger(\hat{a},\theta)\,\right]^{\otimes n} \sigma_{k\alpha}\,\left[U(\hat{a},\theta)\right]^{\otimes n}\right\}\nonumber \\ 
&=&I_2^{\otimes n}\otimes \frac{1}{2}\,\sum_{k=1}^{n}\,\left\{\sum_{\beta=1}^{3}\, R_{\alpha\beta}(\hat{a},\theta)\,  \sigma_{k\beta}\right\} \nonumber \\
&=&\sum_{\beta=1}^3 R_{\alpha\beta}(\hat{a},\theta)\, J_{B\beta}  
\end{eqnarray} 
where $U(\hat{a},\,\theta)=\exp \left(\frac{-i(\sigma\cdot{\hat a})\theta}{2}\right)\in SU(2)$  denotes local unitary operation and     
 $R_{\alpha\beta}(\hat{a},\theta)$, $\alpha,\, \beta=1,\,2,\,3$ denote the elements of the corresponding $3\times 3$ proper orthogonal rotation matrix  $R(\hat a,\, \theta) \in SO(3)$,  with $\hat{a},\ \theta$ being the  axis, angle parameters.

In a symmetric even $N$-qubit state, the angular momentum operators $J_{A\alpha}$, $J'_{B\alpha}$, $\alpha=1,\,2,\,3$  obey the following sum uncertainty relations (see (\ref{main})): 
\begin{eqnarray}
& &\sum_{\alpha=1}^3 (\Delta J_{A\alpha})^2 \geq \frac{n}{2}, \ \ \ \ \  \sum_{\alpha=1}^3 (\Delta J'_{B\alpha})^2\geq \frac{n}{2}, 
\end{eqnarray}
where $n=N/2$. Then the ensuing  LSUR (See (\ref{lsur})) 
\be
\label{lsursym}
\sum_{\alpha=1}^3 \left[\Delta  \left(J_{A\alpha}+ J'_{B\alpha}\right)\right]^2\geq n.
\ee
is satisfied by the set of all separable symmetric states. 

Violation of the LSUR (\ref{lsursym}) i.e.,  
$\sum_{\alpha=1}^3 \left[\Delta  \left(J_{A\alpha}+ J'_{B\alpha}\right)\right]^2~<~n$  reveals that the bipartite state
$\rho^{\rm sym}_{AB}$ is  entangled. 

We define 
\begin{eqnarray}
\label{chin}
&&\chi(\hat{a},\theta)=\frac{1}{n^2}\left( \sum_{\alpha=1}^3\left\{ \Delta \left[ J_{A\alpha}+ J'_{B\alpha}\right]\right\}^2-n\right)  =\frac{1}{n^2}\left(  \sum_{\alpha=1}^3\left\{\Delta \left[J_{A\alpha}+\sum_{\beta=1}^3\, R_{\alpha\beta}(\hat{a},\theta)\,J_{B\beta}\right]\right\}^2-n\right)  
\end{eqnarray}
and observe that  $\chi(\hat{a},\theta)<0 \Longrightarrow$ the LSUR (\ref{lsursym}) is violated. This in turn implies that  $\rho^{\rm sym}_{AB}$ is entangled. 

We now proceed to prove the following Lemma. 

\noindent{\bf Lemma:}
 In an even $N$-qubit permutation symmetric state $\rho_{AB}^{\rm sym}$ it is seen that  
\be
\label{lemF}
\chi(\hat{a},\theta)=\frac{1}{2}\left(\,1- s_0^2+{\rm Tr}\, [R(\hat{a},\theta)\,C]\right)
\ee 
where   $C=T-ss^T$ is the  $3\times 3$ real symmetric {\em covariance matrix}~\cite{pla07,prl07} defined in terms of the two-qubit correlation matrix $T=(t_{\alpha\beta}),\ \alpha,\beta=1,2,3,  {\rm Tr}\,[T]=1$ (see (\ref{tab0})) and the qubit orientation vector $s=(s_1, s_2,s_3)^T$ (see (\ref{salpha0})). The squared magnitude of the qubit orientation vector is denoted by $s_0^2=s^T s=s_1^2+s_2^2+s_3^2$.   

\noindent{\bf Proof:}
  Let us express  
\begin{eqnarray*}
\left[\Delta (J_{A\alpha}+J'_{B\alpha})\right]^2&=&\left[\Delta (J_{A\alpha})\right]^2+\left[\Delta (J'_{B\alpha})\right]^2 + 2\langle 
J_{A\alpha} J'_{B\alpha} \rangle - 2\langle J_{A\alpha}\rangle\, \langle J'_{B\alpha} \rangle 
\end{eqnarray*} 
and simplify the left hand side of the  LSUR (\ref{lsursym}) by substituting  (\ref{s1})--(\ref{scor}), (\ref{salpha0}),  (\ref{tab0}) to obtain  
\begin{eqnarray}
\label{final}
\sum_{\alpha=1}^3\, \left[\Delta (J_{A\alpha}+J'_{B\alpha})\right]^2&=&\frac{n(n+2)}{2}-\frac{n^2}{2}\, s_0^2 +\frac{n^2}{2}\, {\rm Tr}\,
\left[ R(\hat{a},\theta) C\right]. 
\end{eqnarray} 
It readily follows that  $\chi(\hat{a},\theta)$ defined in (\ref{chin}) reduces to the simple form given by (\ref{lemF}).   \hskip 1.3in $\square$

Our main result is presented in the form of the following theorem: \\
\noindent{\bf Theorem:}
	Permutation symmetric even $N$-qubit state violates the angular momentum LSUR (\ref{lsursym}) if the associated two-qubit covariance matrix $C=T-ss^T$ has a strictly negative eigenvalue.    

\noindent{\bf Proof:}
	Violation of the LSUR (\ref{lsursym}) is ensured whenever there exist axis-angle parameters  $a~=~(a_1,a_2,a_3)^T$, $a^Ta=a_1^2+a_2^2+a_3^2=1$, $0\leq \theta\leq 2\pi$ such that $\chi(\hat{a},\theta)<0.$ We show in the following that it is possible to choose these parameters such that $\chi(\hat{a},\theta)<0$ ensuring violation of LSUR (\ref{lsursym}) when the two-qubit covariance matrix $C=T-ss^T$ is not positive-definite possessing atleast one negative eigenvalue. 

Substituting the explicit form~\cite{kns} for the elements of the rotation matrix  $R(\hat{a},\theta)$ i.e.,   
	\begin{eqnarray}
	\label{rot}
	R_{\alpha\beta}&=&\cos \theta\, \delta_{\alpha\beta}+(1-\cos \theta)a_\alpha a_\beta-\sin \theta\,\epsilon_{\alpha\beta\gamma}\,a_\gamma, 
	\hskip 1in \alpha,\,\beta,\,\gamma=1,\,2,\,3
	\end{eqnarray} 
in terms of the axis-angle parameters  and using the property 
$c_{\alpha\beta}=c_{\beta\alpha}$
of the covariance matrix $C=T-ss^T$, 
 we obtain 
\begin{eqnarray}
\label{rotC}
 {\rm Tr}\, [R(\hat{a},\theta)\,C]&=& \cos \theta\, {\rm Tr}\, [C] + (1-\cos \theta)\, a^T\,C\,a = \cos\theta\, (1-s_0^2) + (1-\cos \theta)\, a^TCa.   
 \end{eqnarray}
 where we have substituted  
 \be
 {\rm Tr}\, [C]=(1-s_0^2).
 \ee
 Furthermore, choosing  the rotation axis  $a=(a_1,a_2,a_3)^T$ to be the eigenvector of the two-qubit covariance matrix $C$, corresponding to its least eigenvalue $c_L$,   and setting  the rotation angle to be $\theta=\pi$ in (\ref{rotC}) we get 
  \begin{eqnarray}
  \label{rotCf}
  {\rm Tr}\, [R(\hat{a},\theta=\pi)\,C]&=& -(1-s_0^2) + 2\, c_L.  
  \end{eqnarray}
 Substituting (\ref{rotCf}) in (\ref{lemF}) leads to   
 \begin{equation}
\chi(\hat{a},\theta=\pi)= c_L.  
 \end{equation} 
It is thus evident that the LSUR (\ref{lsursym}) is violated, whenever $c_L<0$,  hence proving the Theorem.  \hskip 0.7in $\square$

In Ref.~\cite{pla07} it has been established by some of us that a two-qubit symmetric state $\varrho^{\rm sym}$ is entangled (negative under partial transpose) if and only if its associated covariance matrix $C=T-ss^T$ is negative. Thus the above Theorem draws attention to the fact that entanglement in the two-qubit reduced state $\varrho^{\rm sym}$ of the whole symmetric even $N$-qubit system $\rho_{AB}^{\rm sym}$ reflects itself in the violation of the LSUR (\ref{lsursym}). 

Substituting (\ref{rotCf}) in (\ref{final}), the left hand side of LSUR, one obtains     
\begin{eqnarray}
\label{lsurlhs}
\sum_{\alpha=1}^3\, \left[\Delta (J_{A\alpha}+J'_{B\alpha})\right]^2=n(1+n\, c_L).  
\end{eqnarray}
Thus 
\be
\label{lsurc}
\sum_{\alpha=1}^3\, \left[\Delta (J_{A\alpha}+J'_{B\alpha})\right]^2\geq n \Rightarrow (1+n\, c_L)\geq 1
\ee 
where  LSUR (\ref{lsursym}) is expressed in terms of the least eigenvalue of the two-qubit covariance matrix $C$. As long as $c_L\geq 0$ (which happens to be the case for  separable symmetric states) the LSUR (\ref{lsurc})  is obeyed. 

We discuss two specific physical examples of $N$-qubit symmetric states which violate (\ref{lsurc}) in the following section.

\section{Examples of $N$-qubit symmetric states violating the angular momentum LSUR }
\subsection{Symmetric multiqubit state generated by one-axis twisting Hamiltonian} 
Kitagawa and Ueda~\cite{ku} had proposed a non-linear Hamiltonian $$\hat{H}=\chi J^2_1,$$ referred to as the {\em   one-axis twisting Hamiltonian} for generating multiqubit spin squeezed states, where $J_1$ denotes one of the components of the collective angular momentum operator of the $N$-qubit system. 

Dynamical evolution of an initially {\em spin-down} state $\vert j,\,-j\rangle$ of the \break $N=2j$ qubit system,  governed by the one-axis twisting  Hamiltonian, results in~\cite{ku}: 
\begin{eqnarray}
\label{oath} 
\vert\Psi_{\rm {KU}}\rangle&=&\exp(-iHt)\vert j,\,-j\rangle, \ \ j=\frac{N}{2}. 
\end{eqnarray}
\begin{figure}[ht]
	\label{1}
	\begin{center}
		\includegraphics*[width=3in,keepaspectratio]{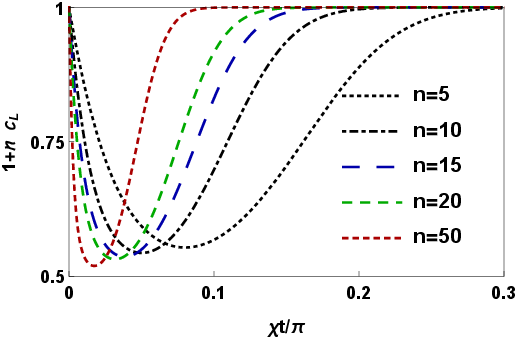}
		\caption{(Color online)  A plot of $(1+n\, c_L)$, the left hand side of the LSUR (\ref{lsurc}), in the  $N$-qubit symmetric state $\vert\Psi_{\rm {KU}}\rangle$ given in (\ref{oath}), as a function of the dimensionless dynamical parameter $\chi t$ for different choices of  $n=N/2$.}
	\end{center}
\end{figure}

In Ref. \cite{ijmp06,aups}, authored by some of us, the mean spin vector $s=(s_1,s_2,s_3)^T$ and the correlation matrix $T$ of a random pair of qubits drawn from the state $\vert\Psi_{\rm {KU}}\rangle$ were explicitly given: The mean spin vector is given by  
\be 
{s}^T=\left(0,\,0,\,-\cos^{(N-1)}(\chi\,t)\right)
\ee    
and the the non-zero elements of the $3\times 3$ real symmetric matrix $T$ take the form, 
\begin{eqnarray}
& & t_{11}=t_{13}=t_{23}=0, \ \  \ t_{12}=\cos^{(N-2)}(\chi\,t)\,\sin(\chi\,t) \ \nonumber\\
& & t_{22}=\frac{1}{2}\left[1- \cos^{(N-2)}(2\chi\,t) \right],\ \  t_{33}=1-t_{22}
\end{eqnarray}   
We then construct the covariance matrix $C=T-{s}\, {{s}}^T$ and evaluate its eigenvalues as a function of the number of qubits $N$ and the dimensionless dynamical parameter $\chi\,t$. Fig.~1 illustrates the behaviour of the left hand side of the LSUR (\ref{lsurc})  with respect to $\chi\,t$ for different choices of $n=N/2$. 

Violation of LSUR (\ref{lsurc}) is clearly seen in Fig. 1, as $1+ n c_L< 1$ and it is a signature of entanglement in the symmetric $N$-qubit state (\ref{oath}), where bipartite divisions are characterized by the collective angular momenta  $j_A=j_B=n/2$.

\subsection{One-parameter family of W-class $N$-qubit states}
We consider the one-parameter   $N$-qubit  symmetric state of the W-class~\cite{adsum}: 
\begin{eqnarray}
\label{wclass}
\vert \Psi\rangle_{\rm W}&=&\, a\,\vert 0_1,0_2,\ldots, 0_N\rangle+\sqrt{1-a^2}\,\vert W\rangle,   \hskip 1in 0<a<1, 
\end{eqnarray}
where
\begin{eqnarray}
\vert W\rangle &=&\frac{1}{\sqrt{N}}\, \left(\vert 1_1,0_2,\ldots\,0_N\rangle + \vert 0_1,1_2,0_3,\ldots\,0_N\rangle +\ldots\ldots +\vert 0_1,0_2,\ldots\,1_N\rangle \right) 
\end{eqnarray}
denotes the symmetric $N$-qubit  W state. 

Reduced two-qubit density matrix $\varrho^{\rm sym}={\rm Tr}_{N-2}\, [\,\vert \Psi\rangle_{\rm W}\langle \Psi\vert\, ]$ obtained by tracing any of the $N-2$ qubits is found to be ~\cite{adsum} 
\be
\label{2wclass}
\varrho^{\rm sym}=\frac{1}{A+2D}\left(\begin{array}{cccc} A & B &B & 0\\ 
	B & D &D & 0\\  B & D &D & 0\\ 
	0 & 0 & 0 & 0        \end{array}\right)
\ee
where    
\begin{eqnarray}
A&=&\frac{N^2a^2+(N-2)(1-a^2)}{N^2\,a^2+N(1-a^2)},\ \ B=\frac{a\sqrt{1-a^2}}{1+a^2(N-1)},\  \ 
D=\frac{1-a^2}{N^2\,a^2+N(1-a^2)}.\ 
\end{eqnarray}
The covariance matrix elements $c_{\alpha\beta}=t_{\alpha\beta}-s_{\alpha}s_{\beta}$ are readily evaluated in the two-qubit symmetric state $\varrho^{\rm sym}={\rm Tr}_{N-2}\, [\,\vert \Psi\rangle_{\rm W}\langle \Psi\vert\, ]$ and  
the associated $3\times 3$ matrix $C=T-ss^T$ takes the form   
\be
\label{cwclass}
C=\frac{1}{(A+2D)^2}\left(\begin{array}{ccc} 2D(A+2D)-4B^2 & 0 & 4BD \\ 
	0 & 2D(A+2D) & 0 \\  4BD & 0 & -4D^2  
\end{array}\right)
\ee
\begin{figure}[h]
	\label{1}
	\begin{center}
		\includegraphics*[width=3in,keepaspectratio]{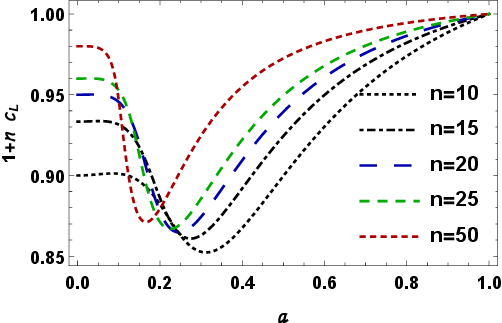}
		\caption{(Color online) Plot of  $(1+n\, c_L)$ i.e., the left hand side (\ref{lsurc}) of the LSUR, in the W-class $N$-qubit symmetric state (\ref{wclass}),   as a function of the parameter $a$ for different values of  $n=N/2$. $c_L$ denotes the minimum eigenvalue of the covariance matrix $C$ (see (\ref{cwclass})) associated with the W-class state.}
	\end{center}
	\end{figure} 

We have plotted  the left hand side of the LSUR (\ref{lsurc}) as a function of the parameter $a$, for different choices of the angular momenta  $j_A=j_B=n/2$ in Fig.~2. Violation of the LSUR i.e., $(1+n\, c_L) \leq 1$ is manifestly seen in Fig.~2, revealing entanglement in the bipartite divisions (characterized by collective angular momenta $j_A=j_B=n/2$) of the W-class $N$-qubit state (\ref{wclass}).

\section{Concluding remarks} 
In this paper, we have shown that angular momentum LSUR in bipartitions of symmetric multiqubit states with even number of qubits are violated when the covariance matrix of the two-qubit subsystem is not positive-definite.  Entanglement between equal bipartitions is ensured when the LSUR in angular momentum operators of an even $N$-qubit symmetric state is violated or equivalently the covariance matrix of the two-qubit subsystem possesses atleast one negative eigenvalue. Our illustration of this result in two important classes of symmetric multiqubit states helps in discerning entanglement in their equal bipartitions through the parameters of two-qubit reduced system.  

One of the main advantages of employing LSUR is the fact that it is possible to detect entanglement without a complete knowledge of the quantum state  and  it suffices to determine experimental friendly  variances of local angular momentum observables for this purpose. 
As violation of LSUR is a collective feature of entanglement in symmetric multiqubit states, it is of interest to contrast this with 
another experimental friendly feature namely, spin squeezing~\cite{xw1,xw2}.  Spin squeezing is only a sufficient collective criterion for pairwise (two-qubit) entanglement in multiqubit symmetric states. Not all entangled symmetric states are spin squeezed.
On the other hand, violation of LSUR reflects itself as a collective feature which is both necessary and sufficient for pairwise entanglement. 
A promising future direction lies in formulating LSUR for higher rank irreducible tensors constructed from angular momentum operators~\cite{multipole} so that their violation brings forth genuine entanglement beyond two-qubit quantum correlations in the global symmetric system. 

\section*{Acknowledgements}
HSK acknowledges the support of NCN through grant SHENG (2018/30/Q/ST2/00625). IR, Sudha and ARU are supported by the Department of Science and Technology, India (Project No. DST/ICPS/QUST/Theme-2/2019).

\end{document}